\documentclass{svproc}
\usepackage{url}

\usepackage{epsfig}  
\usepackage{graphicx}  

\begin{document}
\mainmatter              
\title{Why T2K should run in dominant neutrino mode to discover CP violation ?}
\titlerunning{Why T2K should run in dominant neutrino mode to discover CP violation ?}  
%
\author{Monojit Ghosh}
\authorrunning{Monojit Ghosh} 
%
\institute{Department of Physics, Tokyo Metropolitan University, Hachioji, Tokyo 192-0397, Japan,\\
\email{monojit@tmu.ac.jp}
}

\maketitle              

\begin{abstract}
The first hint of the leptonic CP phase $\delta_{CP}=-90^\circ$ has already came from the long-baseline neutrino oscillation experiment T2K. 
This hint is derived from the neutrino data of T2K and currently it 
is running in the antineutrino mode. In this work we ask the question what should be the proportion of neutrino and antineutrino running of the T2K experiment 
to discover CP violation in the leptonic sector. 

\end{abstract}
\section{Introduction}

Neutrino oscillation in standard three flavour is defined by three mixing angles i.e., $\theta_{12}$, $\theta_{13}$, $\theta_{23}$, two mass squared differences i.e., $\Delta m^2_{21}$, $\Delta m^2_{31}$ and the 
the phase $\delta_{CP}$. Among these six parameters at these moments the unknowns are: (i) neutrino mass hierarchy i.e., normal or inverted (NH: $\Delta m^2_{31} > 0$ or IH: $\Delta m^2_{31} < 0$), 
(ii) the octant of the mixing angle $\theta_{23}$ i.e., lower or higher (LO: $\theta_{23} < 45^\circ$ or HO: $\theta_{23} > 45^\circ$) and the leptonic CP phase $\delta_{CP}$. The first hint of CP violation
in the leptonic sector is believed to come from the currently running long-baseline experiment T2K \cite{Abe:2015awa} in Japan which has already indicated towards a mild preference towards $\delta_{CP}=-90^\circ$. 
This hint has come from the neutrino data of T2K \cite{Abe:2015awa} and currently it is running in the antineutrino mode. 
In this work we ask the question should be the proportion of neutrino and antineutrino run to extract the
best sensitivity from T2K regarding the discovery of leptonic CP violation. 
The capability of T2K to determine the phase
$\delta_{CP}$ is limited by parameter degeneracies \cite{Ghosh:2015ena} which are (i) hierarchy-$\delta_{CP}$ degeneracy and (ii) octant - $\delta_{CP}$ degeneracy.
It is already shown that the the hierarchy - $\delta_{CP}$ degeneracy 
is same for neutrinos and antineutrinos but octant - $\delta_{CP}$ degeneracy behaves differently for 
neutrinos and antineutrinos. Thus a combination of neutrino and antineutrino can resolve the octant - $\delta_{CP}$ degeneracy but not the hierarchy- $\delta_{CP}$ degeneracy. In this work we will show that
for T2K, if the parameter space is free from octant degeneracy then best CP sensitivity comes from the pure neutrino run of T2K. On the other hand antineutrinos are required for the parameter space where
octant degeneracy is present. To overcome this problem we also study the possibility of adding data from other experiments namely the ongoing accelerator base long-baseline experiment NO$\nu$A \cite{Adamson:2016xxw} 
in Fermilab and the proposed atmospheric neutrino experiment ICAL@INO \cite{Ahmed:2015jtv} in India to show that the maximum CP sensitivity of T2K comes from
the dominant neutrino run.

\section{Results and Discussions}

For the simulation of T2K experiment we consider a total exposure of $8 \times 10^{21}$ protons of target (pot). We have divided this exposure in different proportion of neutrino and antineutrino running in units of
$10^{21}$ pot.

\begin{figure}[t!]
\includegraphics[width=0.6\linewidth]{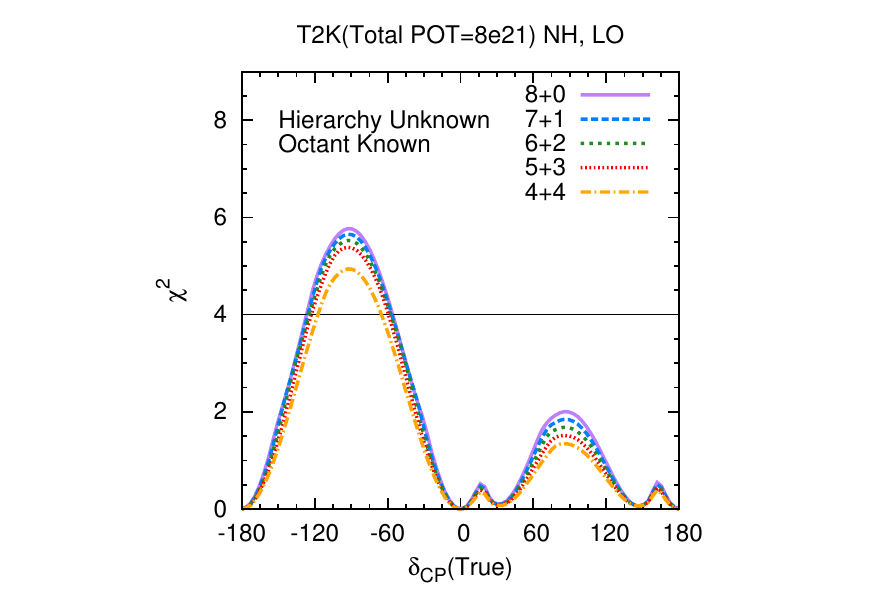}
\hspace{-0.9 in}
\includegraphics[width=0.6\linewidth]{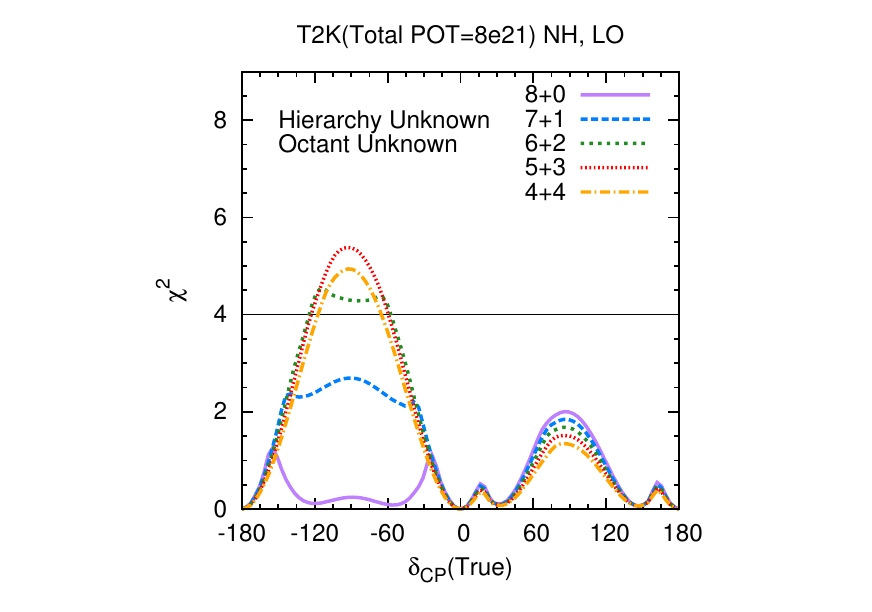} \\
\caption{CP Sensitivity of T2K for NH-LO.}
\label{fig1}
\end{figure}

In Fig. \ref{fig1}, we have plotted the CP violation discovery potential of T2K for NH ($\Delta m^2_{31} = 0.0024$ eV$^2$)-LO ($\theta_{23}=39^\circ$). From  the figure we see that when octant is known (left panel),
the best sensitivity comes from the pure neutrino run i.e., 8+0 configuration. But when octant is unknown (right panel), 8+0 gives the worst sensitivity. As the proportion of antineutrinos increases, CP
sensitivity gets improved. The maximum sensitivity is for 5+3 and further increase of antineutrinos decreases the sensitivity. This is because for 5+3, the wrong-octant solution is completely removed and further addition 
of antineutrinos reduces the statistics and hence the decrease in the sensitivity. 

\begin{figure}[t!]
\includegraphics[width=1.0\linewidth]{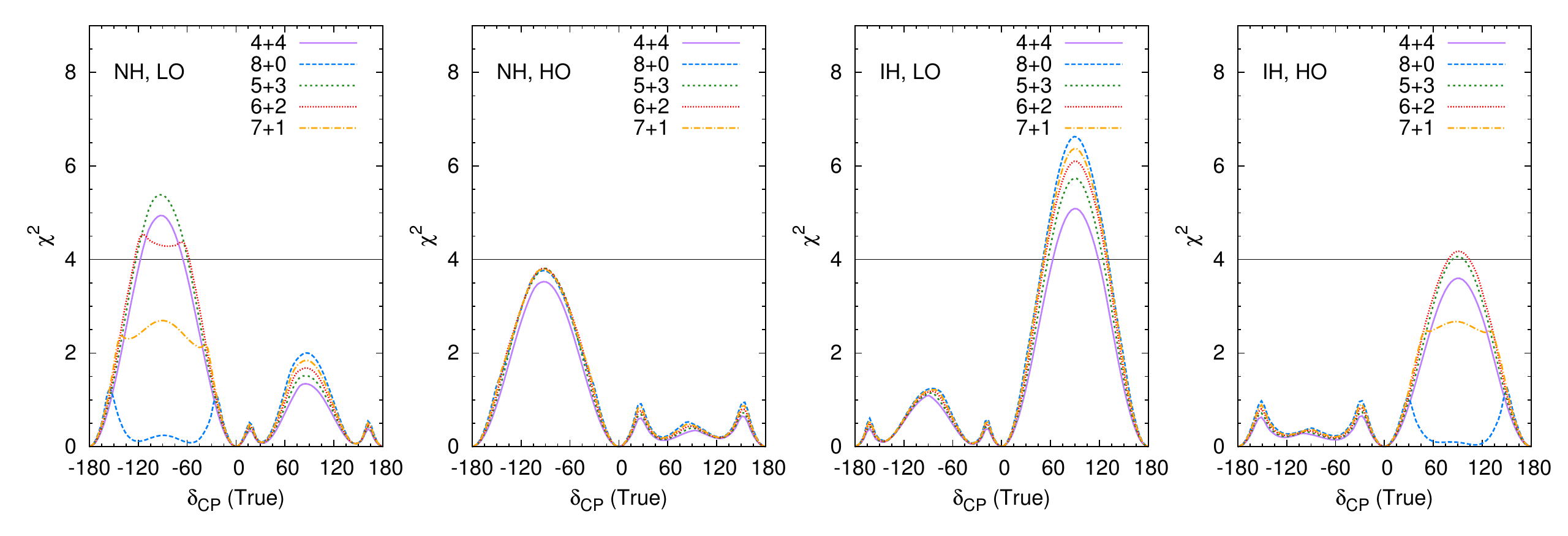}
\caption{CP Sensitivity of T2K for all the four combinations of hierarchy and octant.}
\label{fig2}
\end{figure}
\begin{figure}
\includegraphics[width=1.0\linewidth]{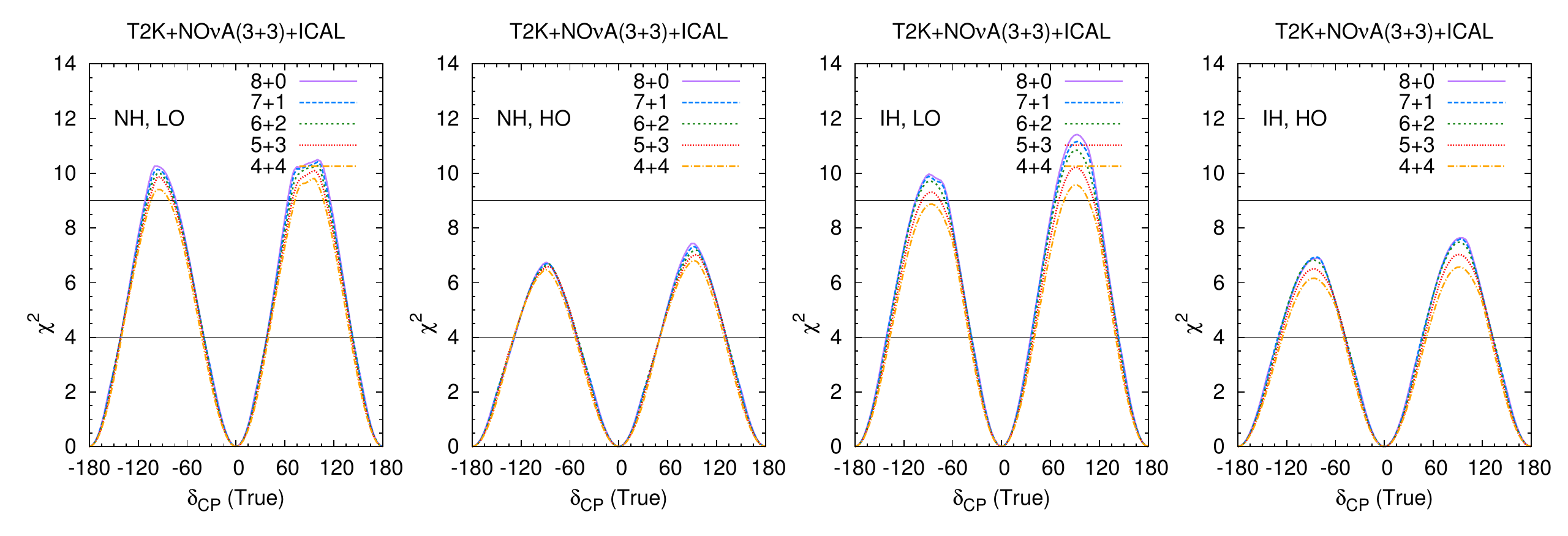}
\caption{CP Sensitivity of T2K, NO$\nu$A and ICAL for all the four combinations of hierarchy and octant.}
\label{fig3}
\end{figure}

In Fig. \ref{fig2} we plotted the same but for all the four combinations of hierarchy and octant assuming octant is unknown. 
IH corresponds to $\Delta m^2_{31} = -0.0024$ eV$^2$ and HO corresponds to $\theta_{23}=51^\circ$. From this figure we see that apart from $-90^\circ$ - NH - LO and $+90^\circ$ - IH - HO, 8+0 configuration of T2K 
gives the best CP sensitivity. Thus to get a handle over these two situations, in Fig. \ref{fig3} we plotted the same as Fig. \ref{fig2} but for the combination of T2K+NO$\nu$A+ICAL.

For NO$\nu$A we assume a three years running in both neutrino and antineutrino mode and for ICAL we consider a 50kt iron calorimeter detector running for 10 years. From the figure we see that when NO$\nu$A and ICAL are combined
with the T2K data then the best CP sensitivity comes from the 7+1 configuration of T2K. 

For further details see Ref. \cite{Ghosh:2015tan} on which this work is based upon.

\section*{Acknowledgements}
This work is partly supported by the Grant-in-Aid for Scientific Research of the Ministry of Education,
Science and Culture, Japan, under Grant No. 25105009.

%
%

\end{document}